\documentclass[twocolumn,xcolor]{aastex631}

\begin{document}

\title{The Near-Infrared Echo from SN 2023xgo: Evidence for a Massive Pre-Supernova Eruption in a Type Ibn/Icn Supernova}

\author[0000-0001-9456-3709]{Masayuki Yamanaka}
\affiliation{Amanogawa Galaxy Astronomy Research Center (AGARC), Graduate School of Science and Engineering, Kagoshima University, 1-21-35 Korimoto, Kagoshima, Kagoshima 890-0065, Japan}

\author{Takahiro Nagayama}
\affiliation{Graduate School of Science and Engineering, Kagoshima University, 1-21-35 Korimoto, Kagoshima, Kagoshima 890-0065, Japan}

\begin{abstract}
We present near-infrared (NIR) and optical observations and analysis of the Type Ibn/Icn supernova (SN) 2023xgo, spanning the period from two days to 100 days past explosion. A comparison of the NIR light curves and color evolution with those of other Type Ibn SNe reveals prominent NIR excess persisting from 15 to 100 days. The spectral energy distribution (SED) is well explained by a combination of a hot blackbody component and a carbon dust model. The dust temperature remained relatively constant at $T=1600\pm100$ K up to $t=60$ d. 
The estimated dust mass is $\sim1.2 \times10^{-4}~M_{\odot}$. 
Based on the SN emission of $10^{42}$ erg s$^{-1}$ at peak luminosity, 
the evaporation radius is estimated to be $1.2\times10^{16}$ cm, 
consistent with the expectation from light travel time.
The dust shell is located outside the shocked cool dense shell (CDS)
at the position of $7.6\times10^{14}$ cm. It suggests that the observed NIR excess originates from pre-existing circumstellar (CS) dust. Considering a typical dust-to-gas mass ratio, the CS gas mass is estimated to be $\simeq10^{-2}~M_{\odot}$, implying a high mass-loss rate of $\sim0.1~M_{\odot}$ yr$^{-1}$ from either a low-mass helium star progenitor with a binary interaction, or may be a massive Wolf-Rayet star that experienced a significant eruption.

\end{abstract}


\section{Introduction} 
 Type Ibn supernovae (SNe~Ibn) are spectroscopically identified by 
 strong narrow helium emission lines and the absence of hydrogen \citep{Foley2007,Pastorello2007c,Pastorello2015a}, indicating presence of dense, helium-rich circumstellar material (CSM).

The progenitor nature of Type Ibn/Icn SNe remains quite ambiguous, even though it is believed that they had undergone multiple episodes of mass-loss within a short time prior to the SN explosion.
The spectra of these SNe show strong, narrow helium emission lines without hydrogen, indicating a stripped progenitor.
The possibility that the progenitor could be a Wolf-Rayet (WR) star in a binary system has been pointed out \citep[][]{Yoon2017, Kool2021, Maeda2022}. Meanwhile, the scenario of a low-mass helium star interacting in a binary system is also suggested \citep{Maund2016, Sun2020}.
 
 Some SNe~Ibn/Icn exhibit a significant near-infrared (NIR) excess in their light curves \citep{Mattila2008,DiCarlo2008,Pastorello2015a,ZYWang2024}. This excess is interpreted as the dust emission, either newly formed within the cool dense shell (CDS) \citep{NSmith2008} or 
 the NIR echo from the pre-existing dust. 
 If the presence of pre-existing dust is confirmed, then the mass-loss rate of the 
 progenitor can be estimated and the constraint on the mass-loss history of the progenitor can be given.
  
SN 2023xgo was discovered by the Zwicky Transient Facility \citep[ZTF; ][]{Bellm2019} at 18.6 mag in the $r$-band on UT 2023 Nov 9
 \citep{Fremling2023}. The spectrum obtained on Nov 12 by \citet{Balcon2023} revealed carbon emission lines superimposed on a blue continuum, leading to its classification as a Type Icn SN. The subsequent follow-up spectroscopy by \citet{Sollerman2023} on Dec 9 detected helium emission lines, indicating a transition to Type Ibn. We defined MJD 60255.6 as $t=0$ d through this Letter based on the polynomial fit to the rising part of the ZTF $r$-band light curve obtained from the ALeRCE site \citep{Forster2021} described in the text. 
  
  The distance to the host galaxy WISEA J050420.81+673723.2 is not reported in the literature.
  The redshift, determined from emission lines in the classification spectrum, 
  is $z=0.01325$ \citep{Balcon2023}, corresponding to a distance of 56~Mpc ($\mu=33.8$ mag) 
  assuming a Hubble constant of $H_{0}=69.8$~km~s$^{-1}$~Mpc$^{-1}$ \citep{Freedman2021}.
  We adopt this distance in this Letter.

   In this Letter, we present the observations and data reduction of SN~2023xgo in Section 2. We show the NIR light curves and color evolution, and an analysis of the spectral energy distribution in Section 3.
   Next, we show the properties of the CS dust and the mass loss rates of the progenitor in Section 4. Finally, we conclude our results in Section 5.
    
\section{Observations and Data Reduction}

\begin{figure}    
   \centering
    \includegraphics[width=0.45\textwidth]{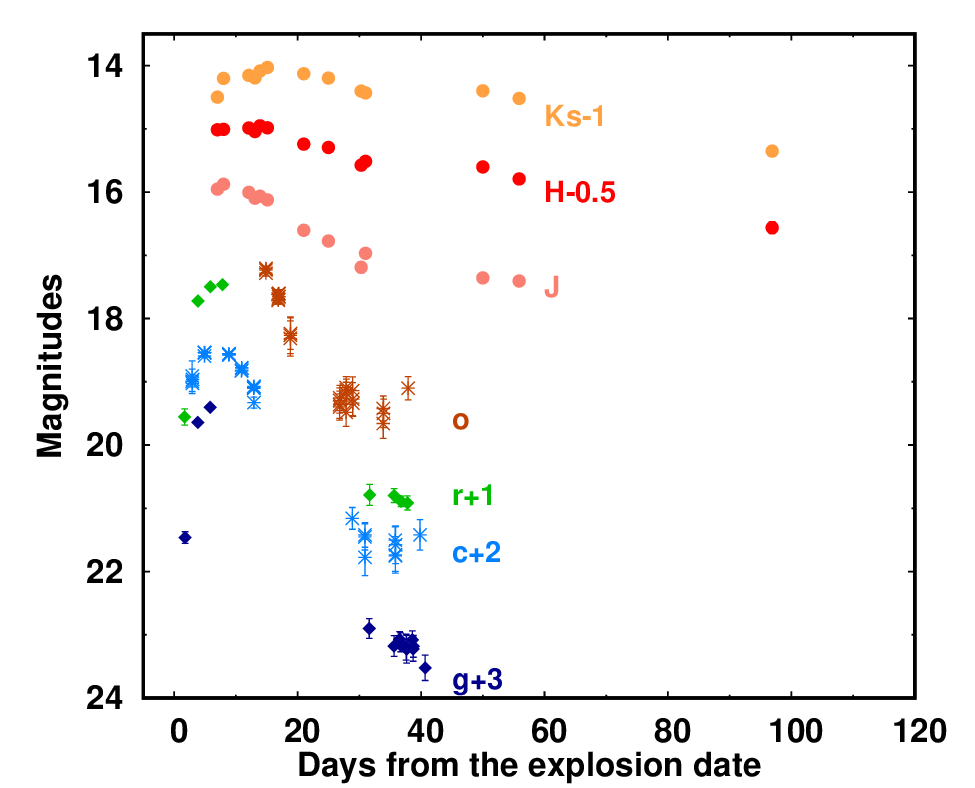}
    \caption{$gcroJHK_{s}$-band light curves of SN 2023xgo. Filled circles represent $JHK_{s}$-band magnitudes from kSIRIUS. Diamonds indicate $g$ and $r$-band magnitudes from ZTF. Asterisks denote $o$ and $c$-band magnitudes from ATLAS. The $JHK_{s}$ magnitudes are in the Vega system, while the $groc$-band magnitudes are in the AB system.}
\end{figure}

\begin{figure}    
   \centering
    \includegraphics[width=0.45\textwidth]{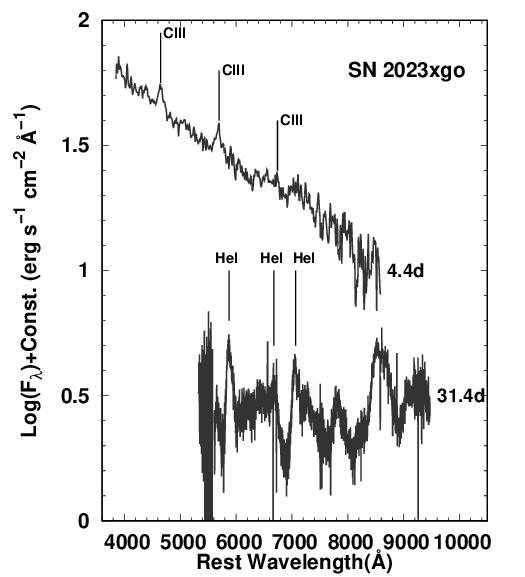}
   \caption{Spectra of SN 2023xgo at $t= 4.4$ and $31.4$ d \citep{Balcon2023,Sollerman2023}. The wavelength has been corrected to its rest wavelength using a redshift of $z=0.01325$. The emission lines detected in the spectra are denoted by the vertical lines.}
\end{figure}
 
  NIR imaging data of SN 2023xgo were obtained on 13 nights from 
  $t=7.0$ to $96.9$ d using the simultaneous three-band NIR imager 
  kSIRIUS \citep{Nagayama2024} attached to the 1-m Kagoshima telescope at the Iriki Observatory. Data reduction was performed using standard IRAF procedures \citep{Tody1986,Tody1993}. Point-spread function (PSF) photometry of the SN 
  and local standard stars was conducted with the DAOPHOT package \citep{Stetson1987}. Photometric calibration was carried out relative to the Two Micron All Sky Survey (2MASS) catalog \citep{Cutri2003}. 
  For the analysis, the $g$- and $r$-band magnitudes 
  obtained through photometry of the template-subtracted images from the ALeRCE system for public ZTF observations were also included \citep{Forster2021}. Additionally, the $orange{\rm (}o{\rm )}$- and $cyan{\rm (}c\rm )$- band magnitudes obtained using the Asteroid Terrestrial-impact Last Alert System \citep[ATLAS;][]{Tonry2018} forced photometry server were also obtained \citep{KWSmith2020,Shingles2021}.
  We plotted the optical and NIR light curves in Figure 1.


 \section{Results}

\subsection{SN typing and interstellar extinction}

 We presented the public spectra obtained by 
 \citet{Balcon2023,Sollerman2023} in Figure 2.
 The carbon emission lines were detected at 4650 \AA\ and 5700 \AA\ with a strong blue continuum at 
 $t=4.4$ d. We measured the full-width at half-maximum (FWHM) of C~{\sc iii}~$\lambda$5696 as 
 $\sim2000$ km~s$^{-1}$. Subsequently, the spectra show helium emission lines. The first classification 
 spectrum showing C~{\sc iii} emission lines was consistent with that of a Type Icn SN, 
 which later evolved into a Type Ibn in the second spectrum 
 \citep[See][for a detailed discussion on the spectral evolution]{Gangopadhyay2025}.

 The Galactic extinction along the line-of-sight to this object was $A_{V}=0.456$ \citep{Schlafly2011}. 
 We did not find any signature of Na~{\sc i} absorption lines in the public spectra (see Figure~2). 
 \cite{Gangopadhyay2025} measured the Na~{\sc i}D absorption and obtained $A_{V}=0.037$ mag for the host galactic extinction. Thus, we applied a total extinction value of $A_{V}=0.493$ mag for the extinction correction throughout this paper.

\subsection{Light curves and color evolution}

 The optical light curves show a rapid rise toward maximum light in the $r$ and $c$ band around $t=8$ d (see Figure 1) at a peak magnitude of $\sim16.5$.
 After reaching peak, $r-$ and $g-$ band light curves experienced a very steep decline until $t=40$ d, with a decline of $\sim7.5$ mag 100~days$^{-1}$ in the $r$ band and 8.8 mag 100~days$^{-1}$ in the $g$ band during this period. These decline rates are significantly faster than 1.5-2.5 mag 100~days$^{-1}$ of Type Ib/c SNe as measured in \citet{Drout2011}.

 The decline rate in $J$, $H$, and $K_{s}$ bands is significantly slower than that of optical bands. In the $J$ band, the decline rate is 1.0 mag per 30 days. In the $H$ and $K_{s}$ 
 bands, the decline rate is 0.5 mag over the same period. 
 SN~2023xgo was detected in the $H$ and $K_{s}$ band data up to 95 days post explosion. The observational data indicate that there is significant NIR emission starting from $t\sim10$ d.

 \begin{figure*}
    \centering
    \includegraphics[width=0.8\textwidth]{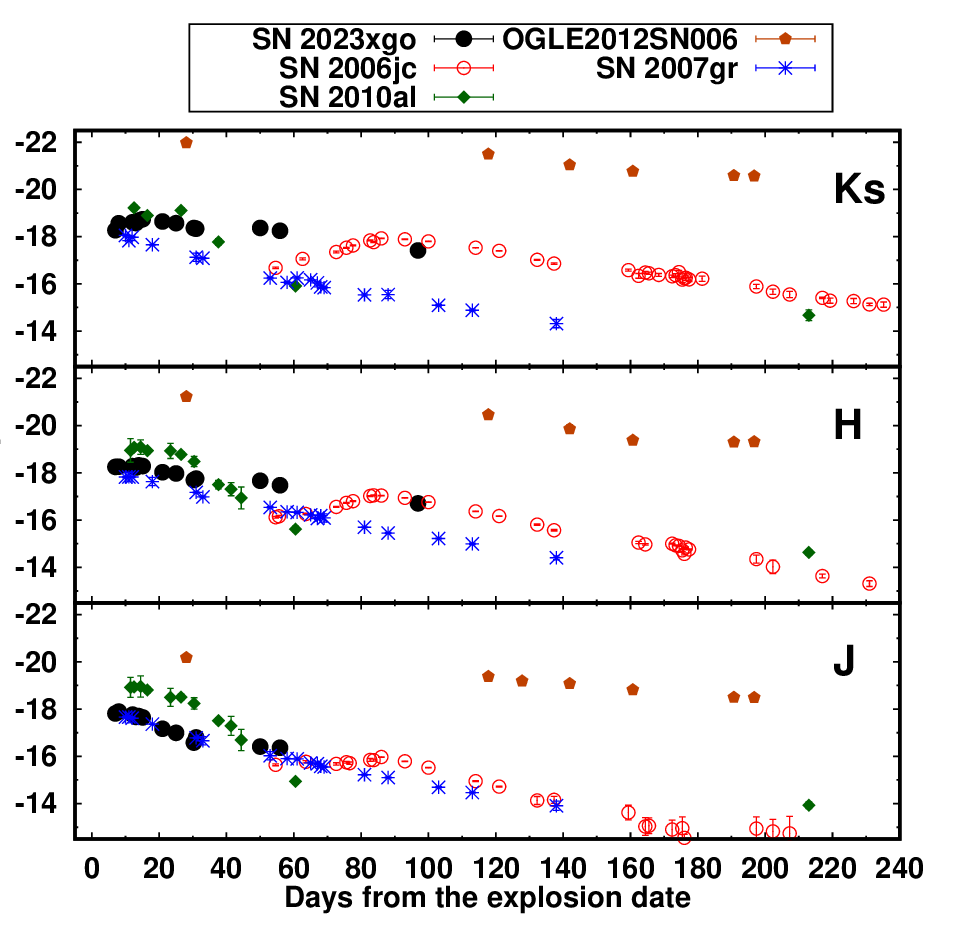}
     \caption{The $JHK_{s}$-band absolute magnitude light curves of SN 2023xgo are 
     presented in each panel, showing the absolute magnitudes in the $J$, $H$, and $K_{s}$ bands. 
     The distance modulus and extinction corrections have already been applied. These light curves are compared with those of Type Ibn SNe 2006jc \citep{Mattila2008,DiCarlo2008}, 2010al \citep{Pastorello2015a}, and OGLE2012SN006 \citep{Pastorello2015b}, as well as a normal SN Ic \citep{Hunter2009}.}
 \end{figure*}

 \begin{figure*}
    \centering
    \includegraphics[width=0.8\textwidth]{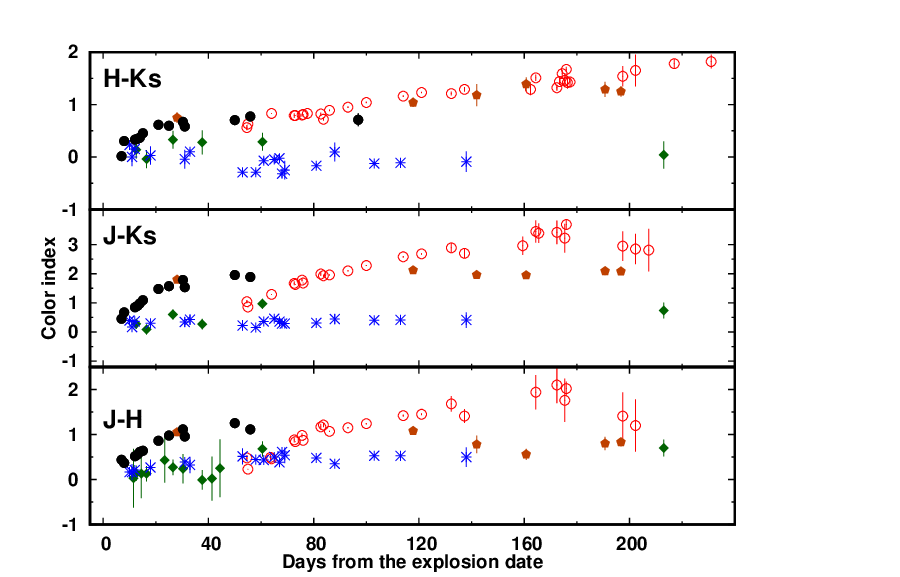}
    \caption{The $H-K_{s}$, $J-K_{s}$, and $J-H$ color evolution of SN 2023xgo is shown, compared with those of SNe 2006jc, 2010al, OGLE2012SN006, and SN 2007gr. 
    The color coding is the same as in Figure 3.}
\end{figure*}

Adopting a distance modulus of $\mu= 33.8 \pm 0.1$ mag for the host galaxy and a total line-of-sight extinction of $A_V = 0.49$ mag (see \S 3.1), we estimate the following absolute magnitudes in the NIR bands: $M_J = -18.0 \pm 0.1$, $M_H = -18.2 \pm 0.1$, and $M_{Ks} = -18.6 \pm 0.1$ mag. The corresponding apparent magnitudes of 15.8, 15.6, and 15.2 mag are estimated from the peaks in their light curves.
 

The light curves were compared with those of other well-studied Type Ibn SNe: 2006jc \citep{Mattila2008}, 2010al \citep{Pastorello2015a}, OGLE2012SN006 \citep{Pastorello2015b}, and Type Ic SN 2007gr \citep{Hunter2009} from $t=0$ to $240$ d.
Within the time span covered by our data (up to $t=60$ d), the $J$-band light curve is very similar to that of SN 2007gr. 
On the other hand, the $H$- and $K_{s}$-band light curves decay more slowly than that in the $J$-band.
On the other hand, the $J$, $H$, and $K_{s}$-band light curves of SN 2006jc show an increase after $t=50$ d. From $t=0$ to $20$ d, SN 2010al is slightly more luminous in absolute magnitude 
compared to SN 2023xgo. In contrast, OGLE2012SN006 is systematically more luminous than SN 2023xgo.

To highlight the onset of the NIR excess, we also present a comparison of the 
NIR color evolution of SN 2023xgo with that of other SNe.
Figure 4 presents the evolution of the $J-H$, $J-K_{s}$, and $H-K_{s}$ 
colors of SN 2023xgo, along with the comparison SNe sample as shown in Figure 3.
The $J-H$ color of SN 2023xgo evolves from approximately 0.5 mag at $t=10$ d to around 1.0 mag at $t=30$
d. Similarly, the $J-K_{s}$ and $H-K_{s}$ colors exhibit the same trend. 
Compared to SN 2006jc, SN 2010al, OGLE2012SN006, and SN 2007gr, the 
$J-H$ color of SN 2023xgo is significantly redder than that of SN 2010al and SN 2007gr. At $t=30$ d, the $J-H$ color is comparable to that of OGLE2012SN006, although the data for SN OGLE2012SN006 is sparse. The $J-H$ color of SN 2023xgo began to evolve towards redder values at $t=10$ d, while that of SN 2006jc evolved at $t=50$ d. Similar trends 
are observed for the $J-K_{s}$ and $H-K_{s}$ colors. 
The $H-K_{s}$ color of SN 2023xgo is comparable to that of
SN 2006jc at $t=90$ d. 



 \section{Discussion}

\subsection{Dust model fitting}

 To explore the origin of the NIR excess, we analyzed the spectral energy distributions (SEDs) with a dust model. First, we constructed 24-epoch SEDs, spanning from $t=7$ to $54$ d. The ZTF $g$-, $r$- and ATLAS $o$-band photometric data were linearly interpolated. These SEDs were created using the central wavelengths of the passband functions and the zero-magnitude fluxes for each band, as defined by \citet{Fukugita1996}, \citet{Tokunaga2005} and \citet{Tonry2018}. 

  Figure 5 shows spectral energy distributions (SEDs) at three representative epochs. We first modeled each SED with a single-temperature blackbody. The model provides an adequate description for $t\lesssim15$ d; at later times the uncertainty in the inferred hot-component temperature increases substantially (Figure 6, top), consistent with a growing NIR flux contribution. The emission lines also contribute to the optical spectra after 
 $t=20$ d \citep[see also][]{Gangopadhyay2025}. The inferred temperature of $T\sim8700\pm1000$ K is consistent with values reported for other SNe Ibn at similar epochs \citep{Mattila2008}. However, it is important to note that this temperature may be underestimated due to the lack of UV flux data. The model fitting cannot be applied to the SED after $t=30$ d because the optical flux decreases.

 After $t=15$ d, we employed a dust model to explain the SED, 
according to the method described in \citet{Fox2010} and \citet{Stritzinger2012}.
The dust component is modeled by the following equation for the SED 
\begin{equation}
F_{\lambda} = B_{\lambda}(T_{d}) \cdot \frac{M_{d} \kappa(\lambda)}{D^{2}}
\end{equation}
where $B_{\lambda}$($T_{d}$) is the Planck function at the dust temperature $T_{d}
$, $M_{d}$ is the dust mass, $\kappa$($\lambda$) is the dust opacity 
(in cm$^2$ g$^{-1}$), and $D$ is the distance to the object (in cm). 
The dust opacity $\kappa$($\lambda$) is further defined as:
\begin{equation}
\kappa(\lambda) = \frac{3Q(a)}{4\rho a}
\end{equation}
Here, $Q$($a$) is the dust emissivity, which depends on the dust grain composition, $\rho$ is the dust density, and $a$ is the typical radius ($0.01$ $\mu$m) of a dust particle \citep{Fox2010,Maeda2013}. For this study, we assume $Q$($a$)$\propto\lambda^{-1.6}$ between 0.4 and 2.5 $\mu$m. and density of 
$\rho=3$ g cm$^{-3}$, assuming a graphite composition \citep{Fox2010}. This assumption is supported by the early classification spectrum obtained 
at $t=4$ d \citep{Balcon2023}, which exhibited carbon emission lines indicative of carbon-rich gas surrounding the progenitor. 

Between \(t=15\) and \(32\)~d, we fit a two-component model to the SEDs, comprising a hot blackbody and a cold dust component. Compared with the single-blackbody fits, the uncertainty in the hot-component temperature decreased substantially, whereas the dust mass remained poorly constrained until \(t\approx 27\)~d. For \(t\gtrsim 32\)~d, we modeled the SEDs with the dust component alone, which reproduces the cold emission well (see Figure~5).

   

\begin{figure}
    \centering
    \includegraphics[width=0.45\textwidth]{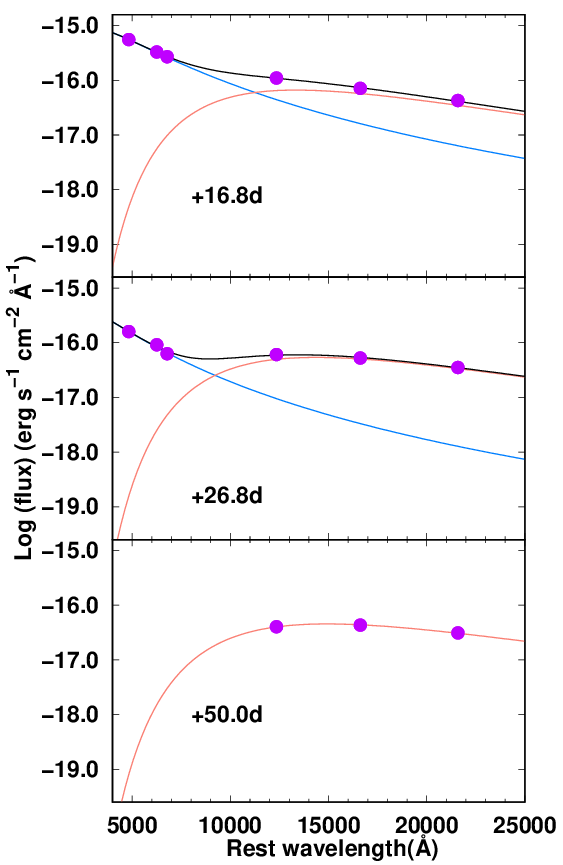}
    \caption{Spectral energy distributions (SEDs) at $t=16.8$, $26.8$, and $50.0$ d. 
    The blue lines show the hot-component blackbody model. The red lines represent 
    the dust model, which explains the NIR excess and is present from $t=16.8$ d. 
   The black lines represent the combined model of the hot component and the dust model.}
\end{figure}

\begin{figure}
    \centering
    \includegraphics[width=0.45\textwidth]{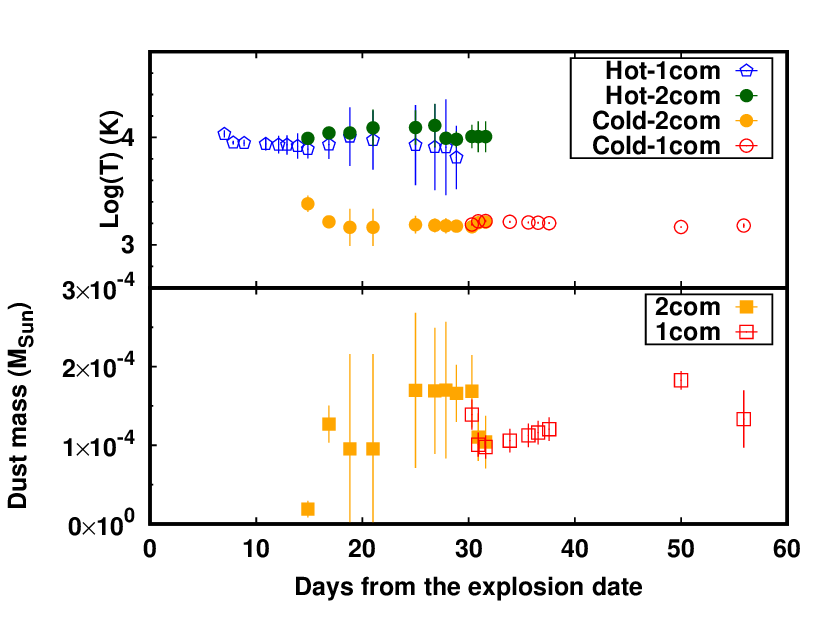}
    \caption{(Top panel) The temporal evolution of the temperature for the hot and cool components of SN 2023xgo. 
    Blue symbols denote the hot component from the single blackbody model. Green symbols denote the hot component, and orange symbols denote the cold component, both from the combined model. Red symbols denote the cold component from the single dust model. 
    Blue, green, and orange symbols represent values obtained using $g$, $r$, $o$, $J$, $H$, and $K_{s}$-band data. 
    Red circles, on the other hand, denote those obtained using $J$, $H$, and $K_{s}$-band data. 
    (Bottom panel) The evolution of the dust mass obtained from the SED fitting. Orange filled squares denote the dust 
    mass obtained using $g$, $r$, $o$, $J$, $H$, and $K_{s}$-band data, while red open squares are obtained using $J$, $H$, and $K_{s}$-band data.}
\end{figure}

The dust component exhibited a relatively constant temperature of $T = 1600 \pm 100$ K throughout the observed 
period ($t = 15$--$55$ d). The estimated dust mass is $(1.2 \pm 0.6) \times 10^{-4} \, M_{\odot}$ for the two-component model between $t=15$ and $32$ d, and $(1.1 \pm 0.4) \times 10^{-4} \, M_{\odot}$ for the single dust model between $t=30$ and $54$ d. 
This relatively constant mass indicates a NIR echo from circumstellar (CS) dust irradiated by the SN radiation. 
The rapid increase in dust mass expected in a dust formation scenario \citep{Kotak2009,Meikle2011,Gall2014} was not observed.


 \subsection{CS dust properties}

Here, we discuss the possibility of carbonaceous dust surviving under the intense SN radiation.
Given the fact that the early-phase SN luminosity is $10^{42}$ erg s$^{-1}$, the dust evaporation radius corresponding to a 
temperature of $T=2000$ K can be estimated to be approximately $1.2\times10^{16}$ cm.
We also obtained the photospheric radius of the hot component to be $7.6 \times 10^{14}$ cm at $t = 15$ d. 
This suggests that the cool dense shell (CDS) region was located significantly interior to the CS dust. Consequently, the NIR excess is likely to originate from pre-existing dust.

Using a light–travel–time argument, the inner radius of the circumstellar (CS) dust is
\(R_{\rm in} \simeq c\,\Delta t\), where \(\Delta t\) is the delay between the illuminating SN light and the onset of the NIR echo. 
Adopting \(\Delta t \simeq 10~\mathrm{d}\) (with the echo first apparent at \(t \simeq 15~\mathrm{d}\)), we obtain 
\(R_{\rm in} \approx c \times 10~\mathrm{d} \approx 2.6 \times 10^{16}\,\mathrm{cm}\), where \(c\) is the speed of light. This agreement supports the conclusion that the CS dust is indeed located at approximately $10^{16}$ cm from the SN.


 Alternatively, if the dust shell has an asymmetric geometry, the actual distance to the CS dust could be larger than the distances estimated from the dust model fitting, because the fitting would only represent an effective emitting area. The asymmetric CSM has been explained before in 2009ip-like SN 2021foa \citep{Gangopadhyay2024}.
 The high degree of polarization detected in some SNe Ibn indicates that the CSM may have an asymmetric distribution \citep{Shivvers2017,Pursiainen2023}.
 Given these findings, the interacting region of SN 2023xgo likely exhibits asymmetry, potentially indicating an asymmetric distribution of the CS dust.

\subsection{Progenitor}

The SED analysis yielded a dust mass estimate of $\sim1.2 \times 10^{-4} \, M_{\odot}$.
Using this value, we discuss the properties of the CSM around the progenitor of SN 2023xgo.
Assuming a standard dust-to-gas mass ratio of approximately 0.01 \citep{Dwek1983}, the corresponding CSM mass is estimated to be on the order of $10^{-2} \, M_{\odot}$.

Analyses of pre-explosion images have suggested that low-mass helium stars stripped off by 
the binary system are potential progenitors of Type Ibn \citep{Maund2016,Sun2020}.
From the public spectra (Figure 2), we estimated from the carbon emission lines that the wind velocity is $\sim2000$ km s$^{-1}$, which is consistent with that of a typical wind velocity of stripped-envelope progenitors. Using this velocity, the estimated mass-loss rate of SN 2023xgo would be approximately $0.1 \, M_{\odot} \, \mathrm{yr}^{-1}$.
The temporal eruption from a low-mass helium star through binary interaction may explain 
this scenario as explained in \citet{Dessart2022}. However, the large wind velocity does not exclude the WR-star progenitor scenario.

 The large mass-loss rate is consistent with the luminous blue variable (LBV) progenitor scenarios transitioning to a WR state. 
 However, the non-detection of hydrogen emission lines is inconsistent with 
 an LBV with a large hydrogen envelope, although a gigantic outburst can 
 reproduce the high-velocity winds \citep{Vink2018,NSmith2020}.
Therefore, based on our analysis, the low-mass helium star from the binary system is a likely progenitor candidate for SN 2023xgo, although we cannot rule single massive stars as the progenitors for SNe Ibn.

 \section{Conclusion}

SN 2023xgo exhibits a significant NIR excess starting $\sim15$ days after the explosion.
The dust model fitting of the SEDs reveals a constant effective temperature of approximately 1600 K.
Assuming a graphite emissivity, we estimated a dust mass of $M_d \sim 1.2\times 10^{-4} \, M_{\odot}$.
The flat evolution of the inferred dust mass from $t=15$ to 55 days suggests a NIR echo from the CS dust.

Based on the peak SN radiation, the carbonaceous dust evaporation radius is estimated to be on the order of $\sim10^{16}$ cm.
The photospheric radius of the shocked gas between the ejecta and the circumstellar material (CSM) is estimated to be $7.6 \times 10^{14}$ cm at early times, indicating that the CS dust is located significantly outside this region.
This location is also consistent with the distance of approximately $2.6 \times 10^{16}$ cm inferred from the light travel time of the NIR echo starting at $t = 15$ days.

With an assumed dust-to-gas mass ratio of approximately 0.01 \citep{Dwek1983}, we estimate the CS gas mass to be approximately $10^{-2} \, M_{\odot}$.
From this gas mass and measuring a wind velocity of 2000 km s$^{-1}$, we estimate a mass-loss rate of approximately $0.1 \, M_{\odot} \, \mathrm{yr}^{-1}$.
This high mass-loss rate implies that the progenitor of SN 2023xgo experienced a substantial mass-loss event, which can be explained by the low-mass stripped-envelope progenitor produced by the binary interaction, or the WR progenitor scenario.


\begin{acknowledgments}
 We are grateful to graduate and undergraduate students for performing 
 the near-infrared observations. We'd like to thank Anjasha Gangopadhyay for their useful comments on the English and scientific input regarding the progenitor scenario. This work was supported by Grant-in-Aid for Scientific Research (C) 22K03676. The Kagoshima University 1 m telescope is a member of the Optical and Infrared Synergetic Telescopes for Education and Research (OISTER) program funded by the MEXT of Japan.
\end{acknowledgments}

\bibliography{addsample}{}

\begin{thebibliography}{}
\expandafter\ifx\csname natexlab\endcsname\relax\def\natexlab#1{#1}\fi
\providecommand{\url}[1]{\href{#1}{#1}}
\providecommand{\dodoi}[1]{doi:~\href{http://doi.org/#1}{\nolinkurl{#1}}}
\providecommand{\doeprint}[1]{\href{http://ascl.net/#1}{\nolinkurl{http://ascl.net/#1}}}
\providecommand{\doarXiv}[1]{\href{https://arxiv.org/abs/#1}{\nolinkurl{https://arxiv.org/abs/#1}}}

\bibitem[{{Balcon}(2023)}]{Balcon2023}
{Balcon}, C. 2023, Transient Name Server Classification Report, 2023-2931, 1

\bibitem[{{Bellm} {et~al.}(2019){Bellm}, {Kulkarni}, {Graham}, {Dekany},
  {Smith}, {Riddle}, {Masci}, {Helou}, {Prince}, {Adams}, {Barbarino},
  {Barlow}, {Bauer}, {Beck}, {Belicki}, {Biswas}, {Blagorodnova}, {Bodewits},
  {Bolin}, {Brinnel}, {Brooke}, {Bue}, {Bulla}, {Burruss}, {Cenko}, {Chang},
  {Connolly}, {Coughlin}, {Cromer}, {Cunningham}, {De}, {Delacroix}, {Desai},
  {Duev}, {Eadie}, {Farnham}, {Feeney}, {Feindt}, {Flynn}, {Franckowiak},
  {Frederick}, {Fremling}, {Gal-Yam}, {Gezari}, {Giomi}, {Goldstein},
  {Golkhou}, {Goobar}, {Groom}, {Hacopians}, {Hale}, {Henning}, {Ho}, {Hover},
  {Howell}, {Hung}, {Huppenkothen}, {Imel}, {Ip}, {Ivezi{\'c}}, {Jackson},
  {Jones}, {Juric}, {Kasliwal}, {Kaspi}, {Kaye}, {Kelley}, {Kowalski},
  {Kramer}, {Kupfer}, {Landry}, {Laher}, {Lee}, {Lin}, {Lin}, {Lunnan},
  {Giomi}, {Mahabal}, {Mao}, {Miller}, {Monkewitz}, {Murphy}, {Ngeow},
  {Nordin}, {Nugent}, {Ofek}, {Patterson}, {Penprase}, {Porter}, {Rauch},
  {Rebbapragada}, {Reiley}, {Rigault}, {Rodriguez}, {van Roestel}, {Rusholme},
  {van Santen}, {Schulze}, {Shupe}, {Singer}, {Soumagnac}, {Stein}, {Surace},
  {Sollerman}, {Szkody}, {Taddia}, {Terek}, {Van Sistine}, {van Velzen},
  {Vestrand}, {Walters}, {Ward}, {Ye}, {Yu}, {Yan}, \& {Zolkower}}]{Bellm2019}
{Bellm}, E.~C., {Kulkarni}, S.~R., {Graham}, M.~J., {et~al.} 2019, \pasp, 131,
  018002, \dodoi{10.1088/1538-3873/aaecbe}

\bibitem[{{Cutri} {et~al.}(2003){Cutri}, {Skrutskie}, {van Dyk}, {Beichman},
  {Carpenter}, {Chester}, {Cambresy}, {Evans}, {Fowler}, {Gizis}, {Howard},
  {Huchra}, {Jarrett}, {Kopan}, {Kirkpatrick}, {Light}, {Marsh}, {McCallon},
  {Schneider}, {Stiening}, {Sykes}, {Weinberg}, {Wheaton}, {Wheelock}, \&
  {Zacarias}}]{Cutri2003}
{Cutri}, R.~M., {Skrutskie}, M.~F., {van Dyk}, S., {et~al.} 2003, VizieR Online
  Data Catalog, II/246

\bibitem[{{Dessart} {et~al.}(2022){Dessart}, {Hillier}, \&
  {Kuncarayakti}}]{Dessart2022}
{Dessart}, L., {Hillier}, D.~J., \& {Kuncarayakti}, H. 2022, \aap, 658, A130,
  \dodoi{10.1051/0004-6361/202142436}

\bibitem[{{Di Carlo} {et~al.}(2008){Di Carlo}, {Corsi}, {Arkharov}, {Massi},
  {Larionov}, {Efimova}, {Dolci}, {Napoleone}, \& {Di Paola}}]{DiCarlo2008}
{Di Carlo}, E., {Corsi}, C., {Arkharov}, A.~A., {et~al.} 2008, \apj, 684, 471,
  \dodoi{10.1086/590051}

\bibitem[{{Drout} {et~al.}(2011){Drout}, {Soderberg}, {Gal-Yam}, {Cenko},
  {Fox}, {Leonard}, {Sand}, {Moon}, {Arcavi}, \& {Green}}]{Drout2011}
{Drout}, M.~R., {Soderberg}, A.~M., {Gal-Yam}, A., {et~al.} 2011, \apj, 741,
  97, \dodoi{10.1088/0004-637X/741/2/97}

\bibitem[{{Dwek}(1983)}]{Dwek1983}
{Dwek}, E. 1983, \apj, 274, 175, \dodoi{10.1086/161435}

\bibitem[{{Foley} {et~al.}(2007){Foley}, {Smith}, {Ganeshalingam}, {Li},
  {Chornock}, \& {Filippenko}}]{Foley2007}
{Foley}, R.~J., {Smith}, N., {Ganeshalingam}, M., {et~al.} 2007, \apjl, 657,
  L105, \dodoi{10.1086/513145}

\bibitem[{{F{\"o}rster} {et~al.}(2021){F{\"o}rster}, {Cabrera-Vives},
  {Castillo-Navarrete}, {Est{\'e}vez}, {S{\'a}nchez-S{\'a}ez}, {Arredondo},
  {Bauer}, {Carrasco-Davis}, {Catelan}, {Elorrieta}, {Eyheramendy}, {Huijse},
  {Pignata}, {Reyes}, {Reyes}, {Rodr{\'\i}guez-Mancini}, {Ruz-Mieres},
  {Valenzuela}, {{\'A}lvarez-Maldonado}, {Astorga}, {Borissova}, {Clocchiatti},
  {De Cicco}, {Donoso-Oliva}, {Hern{\'a}ndez-Garc{\'\i}a}, {Graham},
  {Jord{\'a}n}, {Kurtev}, {Mahabal}, {Maureira}, {Mu{\~n}oz-Arancibia},
  {Molina-Ferreiro}, {Moya}, {Palma}, {P{\'e}rez-Carrasco}, {Protopapas},
  {Romero}, {Sabatini-Gacitua}, {S{\'a}nchez}, {San Mart{\'\i}n},
  {Sep{\'u}lveda-Cobo}, {Vera}, \& {Vergara}}]{Forster2021}
{F{\"o}rster}, F., {Cabrera-Vives}, G., {Castillo-Navarrete}, E., {et~al.}
  2021, \aj, 161, 242, \dodoi{10.3847/1538-3881/abe9bc}

\bibitem[{{Fox} {et~al.}(2010){Fox}, {Chevalier}, {Dwek}, {Skrutskie},
  {Sugerman}, \& {Leisenring}}]{Fox2010}
{Fox}, O.~D., {Chevalier}, R.~A., {Dwek}, E., {et~al.} 2010, \apj, 725, 1768,
  \dodoi{10.1088/0004-637X/725/2/1768}

\bibitem[{{Freedman}(2021)}]{Freedman2021}
{Freedman}, W.~L. 2021, \apj, 919, 16, \dodoi{10.3847/1538-4357/ac0e95}

\bibitem[{{Fremling}(2023)}]{Fremling2023}
{Fremling}, C. 2023, Transient Name Server Discovery Report, 2023-2892, 1

\bibitem[{{Fukugita} {et~al.}(1996){Fukugita}, {Ichikawa}, {Gunn}, {Doi},
  {Shimasaku}, \& {Schneider}}]{Fukugita1996}
{Fukugita}, M., {Ichikawa}, T., {Gunn}, J.~E., {et~al.} 1996, \aj, 111, 1748,
  \dodoi{10.1086/117915}

\bibitem[{{Gall} {et~al.}(2014){Gall}, {Hjorth}, {Watson}, {Dwek}, {Maund},
  {Fox}, {Leloudas}, {Malesani}, \& {Day-Jones}}]{Gall2014}
{Gall}, C., {Hjorth}, J., {Watson}, D., {et~al.} 2014, \nat, 511, 326,
  \dodoi{10.1038/nature13558}

\bibitem[{{Gangopadhyay} {et~al.}(2024){Gangopadhyay}, {Dukiya}, {Moriya},
  {Tanaka}, {Maeda}, {Howell}, {Singh}, {Singh}, {Sollerman}, {Kawabata},
  {Brennan}, {Pellegrino}, {Dastidar}, {Nakaoka}, {Kawabata}, {Misra},
  {Schulze}, {Chandra}, {Taguchi}, {Sahu}, {McCully}, {Bostroem}, {Padilla
  Gonzalez}, {Newsome}, {Hiramatsu}, {Takei}, {Yamanaka}, {Tajitsu}, \&
  {Isogai}}]{Gangopadhyay2024}
{Gangopadhyay}, A., {Dukiya}, N., {Moriya}, T.~J., {et~al.} 2024, arXiv
  e-prints, arXiv:2409.02666, \dodoi{10.48550/arXiv.2409.02666}

\bibitem[{{Gangopadhyay} {et~al.}(2025){Gangopadhyay}, {Sollerman},
  {Tsalapatas}, {Maeda}, {Dukiya}, {Schulze}, {Fransson}, {Sarin}, {Pessi},
  {Singh}, {Wise}, {Nakaoka}, {Singh}, {Dastidar}, {Kawabata}, {Qing}, {Das},
  {Perley}, {Fremling}, {Taguchi}, {Hinds}, {Lunnan}, {Singh Teja}, {Dubey},
  {Ailawadhi}, {Banerjee}, {Kawabata}, {Misra}, {Sahu}, {Brennan}, {Kasliwal},
  {Ho}, {Bochenek}, {Rusholme}, {Laher}, {Smith}, {Purdum}, \&
  {Sravan}}]{Gangopadhyay2025}
{Gangopadhyay}, A., {Sollerman}, J., {Tsalapatas}, K., {et~al.} 2025, arXiv
  e-prints, arXiv:2506.10700, \dodoi{10.48550/arXiv.2506.10700}

\bibitem[{{Hunter} {et~al.}(2009){Hunter}, {Valenti}, {Kotak}, {Meikle},
  {Taubenberger}, {Pastorello}, {Benetti}, {Stanishev}, {Smartt}, {Trundle},
  {Arkharov}, {Bufano}, {Cappellaro}, {di Carlo}, {Dolci}, {Elias-Rosa},
  {Frandsen}, {Fynbo}, {Hopp}, {Larionov}, {Laursen}, {Mazzali}, {Navasardyan},
  {Ries}, {Riffeser}, {Rizzi}, {Tsvetkov}, {Turatto}, \& {Wilke}}]{Hunter2009}
{Hunter}, D.~J., {Valenti}, S., {Kotak}, R., {et~al.} 2009, \aap, 508, 371,
  \dodoi{10.1051/0004-6361/200912896}

\bibitem[{{Kool} {et~al.}(2021){Kool}, {Karamehmetoglu}, {Sollerman},
  {Schulze}, {Lunnan}, {Reynolds}, {Barbarino}, {Bellm}, {De}, {Duev},
  {Fremling}, {Golkhou}, {Graham}, {Green}, {Horesh}, {Kaye}, {Kim}, {Laher},
  {Masci}, {Nordin}, {Perley}, {Phinney}, {Porter}, {Reiley}, {Rodriguez}, {van
  Roestel}, {Rusholme}, {Sharma}, {Sfaradi}, {Soumagnac}, {Taggart},
  {Tartaglia}, {Williams}, \& {Yan}}]{Kool2021}
{Kool}, E.~C., {Karamehmetoglu}, E., {Sollerman}, J., {et~al.} 2021, \aap, 652,
  A136, \dodoi{10.1051/0004-6361/202039137}

\bibitem[{{Kotak} {et~al.}(2009){Kotak}, {Meikle}, {Farrah}, {Gerardy},
  {Foley}, {Van Dyk}, {Fransson}, {Lundqvist}, {Sollerman}, {Fesen},
  {Filippenko}, {Mattila}, {Silverman}, {Andersen}, {H{\"o}flich}, {Pozzo}, \&
  {Wheeler}}]{Kotak2009}
{Kotak}, R., {Meikle}, W.~P.~S., {Farrah}, D., {et~al.} 2009, \apj, 704, 306,
  \dodoi{10.1088/0004-637X/704/1/306}

\bibitem[{{Maeda} \& {Moriya}(2022)}]{Maeda2022}
{Maeda}, K., \& {Moriya}, T.~J. 2022, \apj, 927, 25,
  \dodoi{10.3847/1538-4357/ac4672}

\bibitem[{{Maeda} {et~al.}(2013){Maeda}, {Nozawa}, {Sahu}, {Minowa},
  {Motohara}, {Ueno}, {Folatelli}, {Pyo}, {Kitagawa}, {Kawabata}, {Anupama},
  {Kozasa}, {Moriya}, {Yamanaka}, {Nomoto}, {Bersten}, {Quimby}, \&
  {Iye}}]{Maeda2013}
{Maeda}, K., {Nozawa}, T., {Sahu}, D.~K., {et~al.} 2013, \apj, 776, 5,
  \dodoi{10.1088/0004-637X/776/1/5}

\bibitem[{{Mattila} {et~al.}(2008){Mattila}, {Meikle}, {Lundqvist},
  {Pastorello}, {Kotak}, {Eldridge}, {Smartt}, {Adamson}, {Gerardy}, {Rizzi},
  {Stephens}, \& {van Dyk}}]{Mattila2008}
{Mattila}, S., {Meikle}, W.~P.~S., {Lundqvist}, P., {et~al.} 2008, \mnras, 389,
  141, \dodoi{10.1111/j.1365-2966.2008.13516.x}

\bibitem[{{Maund} \& {Ramirez-Ruiz}(2016)}]{Maund2016}
{Maund}, J.~R., \& {Ramirez-Ruiz}, E. 2016, \mnras, 456, 3175,
  \dodoi{10.1093/mnras/stv2760}

\bibitem[{{Meikle} {et~al.}(2011){Meikle}, {Kotak}, {Farrah}, {Mattila}, {Van
  Dyk}, {Andersen}, {Fesen}, {Filippenko}, {Foley}, {Fransson}, {Gerardy},
  {H{\"o}flich}, {Lundqvist}, {Pozzo}, {Sollerman}, \& {Wheeler}}]{Meikle2011}
{Meikle}, W.~P.~S., {Kotak}, R., {Farrah}, D., {et~al.} 2011, \apj, 732, 109,
  \dodoi{10.1088/0004-637X/732/2/109}

\bibitem[{{Nagayama} \& {Nakaya}(2024)}]{Nagayama2024}
{Nagayama}, T., \& {Nakaya}, H. 2024, in Society of Photo-Optical
  Instrumentation Engineers (SPIE) Conference Series, Vol. 13096, Ground-based
  and Airborne Instrumentation for Astronomy X, ed. J.~J. {Bryant},
  K.~{Motohara}, \& J.~R.~D. {Vernet}, 130963I, \dodoi{10.1117/12.3016593}

\bibitem[{{Pastorello} {et~al.}(2007){Pastorello}, {Smartt}, {Mattila},
  {Eldridge}, {Young}, {Itagaki}, {Yamaoka}, {Navasardyan}, {Valenti}, {Patat},
  {Agnoletto}, {Augusteijn}, {Benetti}, {Cappellaro}, {Boles}, {Bonnet-Bidaud},
  {Botticella}, {Bufano}, {Cao}, {Deng}, {Dennefeld}, {Elias-Rosa},
  {Harutyunyan}, {Keenan}, {Iijima}, {Lorenzi}, {Mazzali}, {Meng}, {Nakano},
  {Nielsen}, {Smoker}, {Stanishev}, {Turatto}, {Xu}, \&
  {Zampieri}}]{Pastorello2007c}
{Pastorello}, A., {Smartt}, S.~J., {Mattila}, S., {et~al.} 2007, \nat, 447,
  829, \dodoi{10.1038/nature05825}

\bibitem[{{Pastorello} {et~al.}(2015{\natexlab{a}}){Pastorello}, {Benetti},
  {Brown}, {Tsvetkov}, {Inserra}, {Taubenberger}, {Tomasella}, {Fraser},
  {Rich}, {Botticella}, {Bufano}, {Cappellaro}, {Ergon}, {Gorbovskoy},
  {Harutyunyan}, {Huang}, {Kotak}, {Lipunov}, {Magill}, {Miluzio}, {Morrell},
  {Ochner}, {Smartt}, {Sollerman}, {Spiro}, {Stritzinger}, {Turatto},
  {Valenti}, {Wang}, {Wright}, {Yurkov}, {Zampieri}, \&
  {Zhang}}]{Pastorello2015a}
{Pastorello}, A., {Benetti}, S., {Brown}, P.~J., {et~al.} 2015{\natexlab{a}},
  \mnras, 449, 1921, \dodoi{10.1093/mnras/stu2745}

\bibitem[{{Pastorello} {et~al.}(2015{\natexlab{b}}){Pastorello}, {Wyrzykowski},
  {Valenti}, {Prieto}, {Koz{\l}owski}, {Udalski}, {Elias-Rosa},
  {Morales-Garoffolo}, {Anderson}, {Benetti}, {Bersten}, {Botticella},
  {Cappellaro}, {Fasano}, {Fraser}, {Gal-Yam}, {Gillone}, {Graham}, {Greiner},
  {Hachinger}, {Howell}, {Inserra}, {Parrent}, {Rau}, {Schulze}, {Smartt},
  {Smith}, {Turatto}, {Yaron}, {Young}, {Kubiak}, {Szyma{\'n}ski},
  {Pietrzy{\'n}ski}, {Soszy{\'n}ski}, {Ulaczyk}, {Poleski}, {Pietrukowicz},
  {Skowron}, \& {Mr{\'o}z}}]{Pastorello2015b}
{Pastorello}, A., {Wyrzykowski}, {\L}., {Valenti}, S., {et~al.}
  2015{\natexlab{b}}, \mnras, 449, 1941, \dodoi{10.1093/mnras/stu2621}

\bibitem[{{Pursiainen} {et~al.}(2023){Pursiainen}, {Leloudas}, {Schulze},
  {Charalampopoulos}, {Angus}, {Anderson}, {Bauer}, {Chen}, {Galbany},
  {Gromadzki}, {Guti{\'e}rrez}, {Inserra}, {Lyman}, {M{\"u}ller-Bravo},
  {Nicholl}, {Smartt}, {Tartaglia}, {Wiseman}, \& {Young}}]{Pursiainen2023}
{Pursiainen}, M., {Leloudas}, G., {Schulze}, S., {et~al.} 2023, \apjl, 959,
  L10, \dodoi{10.3847/2041-8213/ad103d}

\bibitem[{{Schlafly} \& {Finkbeiner}(2011)}]{Schlafly2011}
{Schlafly}, E.~F., \& {Finkbeiner}, D.~P. 2011, \apj, 737, 103,
  \dodoi{10.1088/0004-637X/737/2/103}

\bibitem[{{Shingles} {et~al.}(2021){Shingles}, {Smith}, {Young}, {Smartt},
  {Tonry}, {Denneau}, {Heinze}, {Weiland}, {Flewelling}, {Stalder},
  {Clocchiatti}, {F{\"o}rster}, {Pignata}, {Rest}, {Anderson}, {Stubbs}, \&
  {Erasmus}}]{Shingles2021}
{Shingles}, L., {Smith}, K.~W., {Young}, D.~R., {et~al.} 2021, Transient Name
  Server AstroNote, 7, 1

\bibitem[{{Shivvers} {et~al.}(2017){Shivvers}, {Zheng}, {Van Dyk}, {Mauerhan},
  {Filippenko}, {Smith}, {Foley}, {Mazzali}, {Kamble}, {Kilpatrick},
  {Margutti}, {Yuk}, {Graham}, {Kelly}, {Andrews}, {Matheson}, {Wood-Vasey},
  {Ponder}, {Brown}, {Chevalier}, {Milisavljevic}, {Drout}, {Parrent},
  {Soderberg}, {Ashall}, {Piascik}, \& {Prentice}}]{Shivvers2017}
{Shivvers}, I., {Zheng}, W., {Van Dyk}, S.~D., {et~al.} 2017, \mnras, 471,
  4381, \dodoi{10.1093/mnras/stx1885}

\bibitem[{{Smith} {et~al.}(2020{\natexlab{a}}){Smith}, {Smartt}, {Young},
  {Tonry}, {Denneau}, {Flewelling}, {Heinze}, {Weiland}, {Stalder}, {Rest},
  {Stubbs}, {Anderson}, {Chen}, {Clark}, {Do}, {F{\"o}rster}, {Fulton},
  {Gillanders}, {McBrien}, {O'Neill}, {Srivastav}, \& {Wright}}]{KWSmith2020}
{Smith}, K.~W., {Smartt}, S.~J., {Young}, D.~R., {et~al.} 2020{\natexlab{a}},
  \pasp, 132, 085002, \dodoi{10.1088/1538-3873/ab936e}

\bibitem[{{Smith} {et~al.}(2008){Smith}, {Foley}, \& {Filippenko}}]{NSmith2008}
{Smith}, N., {Foley}, R.~J., \& {Filippenko}, A.~V. 2008, \apj, 680, 568,
  \dodoi{10.1086/587860}

\bibitem[{{Smith} {et~al.}(2020{\natexlab{b}}){Smith}, {E Andrews}, {Moe},
  {Milne}, {Bilinski}, {Kilpatrick}, {Fong}, {Badenes}, {Filippenko},
  {Kasliwal}, \& {Silverman}}]{NSmith2020}
{Smith}, N., {E Andrews}, J., {Moe}, M., {et~al.} 2020{\natexlab{b}}, \mnras,
  492, 5897, \dodoi{10.1093/mnras/staa061}

\bibitem[{{Sollerman} {et~al.}(2023){Sollerman}, {Chu}, {Dahiwale}, \&
  {Fremling}}]{Sollerman2023}
{Sollerman}, J., {Chu}, M., {Dahiwale}, A., \& {Fremling}, C. 2023, Transient
  Name Server Classification Report, 2023-3181, 1

\bibitem[{{Stetson}(1987)}]{Stetson1987}
{Stetson}, P.~B. 1987, \pasp, 99, 191, \dodoi{10.1086/131977}

\bibitem[{{Stritzinger} {et~al.}(2012){Stritzinger}, {Taddia}, {Fransson},
  {Fox}, {Morrell}, {Phillips}, {Sollerman}, {Anderson}, {Boldt}, {Brown},
  {Campillay}, {Castellon}, {Contreras}, {Folatelli}, {Habergham}, {Hamuy},
  {Hjorth}, {James}, {Krzeminski}, {Mattila}, {Persson}, \&
  {Roth}}]{Stritzinger2012}
{Stritzinger}, M., {Taddia}, F., {Fransson}, C., {et~al.} 2012, \apj, 756, 173,
  \dodoi{10.1088/0004-637X/756/2/173}

\bibitem[{{Sun} {et~al.}(2020){Sun}, {Maund}, {Hirai}, {Crowther}, \&
  {Podsiadlowski}}]{Sun2020}
{Sun}, N.-C., {Maund}, J.~R., {Hirai}, R., {Crowther}, P.~A., \&
  {Podsiadlowski}, P. 2020, \mnras, 491, 6000, \dodoi{10.1093/mnras/stz3431}

\bibitem[{{Tody}(1986)}]{Tody1986}
{Tody}, D. 1986, in Society of Photo-Optical Instrumentation Engineers (SPIE)
  Conference Series, Vol. 627, Instrumentation in astronomy VI, ed. D.~L.
  {Crawford}, 733, \dodoi{10.1117/12.968154}

\bibitem[{{Tody}(1993)}]{Tody1993}
{Tody}, D. 1993, in Astronomical Society of the Pacific Conference Series,
  Vol.~52, Astronomical Data Analysis Software and Systems II, ed. R.~J.
  {Hanisch}, R.~J.~V. {Brissenden}, \& J.~{Barnes}, 173

\bibitem[{{Tokunaga} \& {Vacca}(2005)}]{Tokunaga2005}
{Tokunaga}, A.~T., \& {Vacca}, W.~D. 2005, \pasp, 117, 421,
  \dodoi{10.1086/429382}

\bibitem[{{Tonry} {et~al.}(2018){Tonry}, {Denneau}, {Heinze}, {Stalder},
  {Smith}, {Smartt}, {Stubbs}, {Weiland}, \& {Rest}}]{Tonry2018}
{Tonry}, J.~L., {Denneau}, L., {Heinze}, A.~N., {et~al.} 2018, \pasp, 130,
  064505, \dodoi{10.1088/1538-3873/aabadf}

\bibitem[{{Vink}(2018)}]{Vink2018}
{Vink}, J.~S. 2018, \aap, 619, A54, \dodoi{10.1051/0004-6361/201833352}

\bibitem[{{Wang} {et~al.}(2024){Wang}, {Pastorello}, {Maeda}, {Reguitti},
  {Cai}, {Andrew Howell}, {Benetti}, {Buckley}, {Cappellaro}, {Carini},
  {Cartier}, {Chen}, {Elias-Rosa}, {Fang}, {Gal-Yam}, {Gangopadhyay},
  {Gromadzki}, {Gan}, {Hiramatsu}, {Hu}, {Inserra}, {McCully}, {Nicholl},
  {Olivares E.}, {Pignata}, {Pineda-Garc{\'\i}a}, {Pursiainen}, {Ragosta},
  {Rau}, {Roy}, {Sollerman}, {Tartaglia}, {Terreran}, {Valerin}, {Wang},
  {Wang}, {Young}, {Aryan}, {Bronikowski}, {Concepcion}, {Galbany}, {Lin},
  {Melandri}, {Petrushevska}, {Ramirez}, {Shi}, {Warwick}, {Zhang}, {Wang},
  {Wang}, \& {Zhu}}]{ZYWang2024}
{Wang}, Z.~Y., {Pastorello}, A., {Maeda}, K., {et~al.} 2024, \aap, 691, A156,
  \dodoi{10.1051/0004-6361/202451131}

\bibitem[{{Yoon}(2017)}]{Yoon2017}
{Yoon}, S.-C. 2017, \mnras, 470, 3970, \dodoi{10.1093/mnras/stx1496}

\end{thebibliography}
\bibliographystyle{aasjournal}

\end{document}